\documentclass[twocolumn,showpacs,preprintnumbers,amsmath,amssymb]{revtex4}
\usepackage{epsfig}
\begin{document}

\title{Spin transfer in bilayer magnetic nanopillars at high fields \\ as a function of free-layer thickness}

\author{W. Chen$^1$, M. J. Rooks$^2$, N. Ruiz$^2$, J. Z. Sun$^2$, A. D. Kent$^1$}
\affiliation{$^1$Department of Physics, New York University, New
York, NY 10003, USA} \affiliation{$^2$IBM T. J. Watson Research
Center, P.O. Box 218, Yorktown Heights, NY 10598, USA}
\date{August 21, 2006}

\begin{abstract}
Spin transfer in asymmetric Co/Cu/Co bilayer magnetic nanopillars
junctions has been studied at low temperature as a function of
free-layer thickness. The phase diagram for current-induced magnetic
excitations has been determined for magnetic fields up to $7.5 ~T$
applied perpendicular to the junction surface and free-layers
thicknesses from $2$ to $5$ nm. The junction magnetoresistance is
independent of thickness. The critical current for magnetic
excitations decreases linearly with decreasing free-layer thickness,
but extrapolates to a finite critical current in the limit of zero
thickness. The limiting current is in quantitative agreement with
that expected due to a spin-pumping contribution to the
magnetization damping. It may also be indicative of a decrease in
the spin-transfer torque efficiency in ultrathin magnetic
layers.\end{abstract} \maketitle

Spin transfer in magnetic nanopillar has become a major focus of
experimental research \cite{Katine2000, Ozyilmaz2003, Urazhdin2004}
since Slonczewski and Berger's seminal theoretical work in 1996
\cite{Slonczewski1996, Berger1996}. A spin current has been
demonstrated to switch the magnetization direction of a small magnet
at a specific current density, as well as to induce microwave
excitations. There are applications of this effect to magnetic
random access memory (MRAM) and high-frequency electronics
\cite{Katine2000,Kiselev2003, Rippard2004}. It is of importance to
determine the factors that control the critical current for
magnetization dynamics for both the physics and technology of spin
transfer. For instance, it is of interest to reduce the critical
current for MRAM applications, and to increase it in magnetic sensor
designs.

In Slonczewski's theory, spin transfer is an interface effect:
spin-angular momentum is transferred to the background magnetization
when the spin current enters the ferromagnet --within the first few
atomic layers \cite{Stiles2002}. For one polarity of the current,
this generates a torque on the magnetization that is opposed by bulk
damping. As a result, there is a threshold current to excite
magnetization dynamics that is proportional to the volume of the
magnet or, equivalently, the threshold current density is
proportional to the thickness of the magnetic layer. There are
alternative models in which the spin-transfer interaction occurs on
a longer length scale, which predict a decrease in the efficiency of
the torque in very thin magnetic layers \cite{Zhang2004, Zhang2005}.
It is also now widely appreciated that the magnetization damping in
thin layers can be dominated by interfaces, in an effect known as
``spin pumping" \cite{Tserkovnyak2005}. For these reasons it is of
importance to study spin transfer in samples in which the layer
thicknesses are varied to gain insight into the factors that
determine the strength and length scales of the spin-transfer
interaction.

Albert \textit{et al.} \cite{Albert2002} studied current-induced
magnetization switching as a function of free-layer thickness at
room temperature. Here thermal fluctuations are important and the
intrinsic (zero temperature) critical current was determined by
extrapolating from pulsed current measurements. The switching was
between in-plane magnetized states, parallel and antiparallel to the
fixed-layer magnetization. In this case, the in-plane shape
anisotropy plays an important role in setting the energy barrier to
reversal. The switching current was found to depend linearly on the
free-layer thickness and to be zero in the limit of zero free-layer
thickness, consistent with Slonczewski's model.

In this paper, we present studies of spin-transfer at low
temperature and high magnetic fields in asymmetric magnetic
nanopillars in which the free magnetic-layer thickness has been
systematically varied. The phase diagram for magnetic excitations
has been determined in fields perpendicular to the film plane, under
which the in-plane anisotropy is a minor effect, and at low
temperature (4.2 K), where thermal fluctuations can be neglected.
Similar to the results of Albert \textit{et al.} \cite{Albert2002},
we find that the critical current density, defined to be the current
density at which there is a step change in junction resistance, is a
linear function of the layer thickness. However, it extrapolates to
a finite zero thickness intercept. This suggests that damping
related to spin pumping sets a lower limit for the critical current
density in ultra-thin magnetic layers, or that spin transfer occurs
over a finite-length scale in the ferromagnet.

In magnetic nanopillars that consist of thick and thin magnetic
layers separated by a nonmagnetic layer, the thick (fixed) layer
polarizes the current and dynamics is induced in the thin (free)
layer. The magnetization of the free layer can be described by the
Landau-Lifshitz-Gilbert (LLG) equation with an additional
spin-transfer torque term \cite{Slonczewski1996, Berger1996}. In a
macrospin model, which assumes that the two layers are uniformly
magnetized:
\begin{equation}
{{d\hat{m}}\over{dt}}=-\gamma \hat{m} \times H_{\text{eff}} +\alpha
\hat{m} \times {{d\hat{m}}\over{dt}} \\ + {{\gamma a_{J}}\over
{1+\eta \hat{m} \cdot \hat{m_{P}}}} \hat{m}\times (\hat{m}\times
\hat{m_{P}}). \; \label{eq1}
\end{equation}
$\hat{m}$ and $\hat{m_p}$ are unit vectors in the direction of
magnetization of the free and fixed magnetic layers, respectively.
$\gamma$ is the gyromagnetic ratio. The second term on the right is
the damping term, where $\alpha$ is the Gilbert damping constant.
The last term is due to spin-transfer. Here $a_{J}={{\hbar
JP_o}\over{2eM_{\text {s}}t}}$, where $J$ is the current density,
$P_o$ is the spin-polarization of the current, $M_{\text {s}}$ is
the magnetization density, and $t$ is the free magnetic-layer
thickness. The positive constant $\eta$ characterizes the angular
dependence of the torque and depends on the spin polarization
\cite{LiHeZhang2005}. The effective magnetic field $H_{\text {eff}}$
is the vector sum of the applied field $H$, the in-plane anisotropy
field $H_{\text {a}}$, and the easy-plane anisotropy $-4 \pi
M_{\text {eff}}m_{z}\hat{z}$,
$H_{\text{eff}}=H+H_{a}m_{x}\hat{x}-4\pi
M_{\text{eff}}m_{z}\hat{z}$. In equilibrium, the magnetization is
aligned with the effective magnetic field. The spin-transfer torque
competes with damping, and at a threshold current density leads to
excitation of the magnetization.

\begin{figure}[t]
\includegraphics[width=0.48\textwidth]{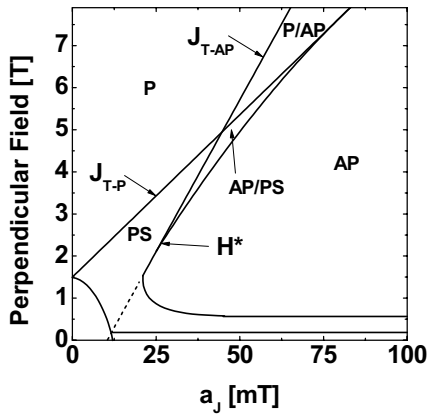}
\caption{Phase diagram of a nanopillar for $\eta=0.3$,
$\alpha=0.01$, and  $4 \pi M_{\text{eff}}=1.5 ~T$, with a small
($0.1~T$) in-plane uniaxial anisotropy. The threshold current
densities for instability of the P state, $J_{T-P}$, and for
instability of the AP state, $J_{T-AP}$, are indicated.}
\label{Numerical}
\end{figure}

A phase diagram assuming a single-domain macrospin model can be
calculated from Eq. 1, as shown in Fig. \ref{Numerical}. For large
fields perpendicular to the plane of the magnetic layers ($H>4\pi
M_{\text {s}}$), the magnetizations of the fixed and free layer are
aligned in the field direction. This is a particularly simple
situation \cite{Ozyilmaz2003}, as the in-plane shape anisotropy of
the element plays only a minor role in the dynamics --the easy-plane
anisotropy dominates, and the magnetic energy has axial symmetry. In
this case, as $J$ increases, the parallel (P) state becomes unstable
at a threshold current:
\begin{eqnarray}
J_{T-P}={2e \alpha \over \hbar P_o}(1+ \eta)M_{\text {s}}t(H - 4\pi
M_{\text {eff}}), \; \label{eq2}
\end{eqnarray}
leading to a precessional (PS) state. A further increase in the
current results in the free layer switching into an antiparallel
(AP) state. For a decreasing current, the AP becomes unstable and
switches back to a PS or P state when:
\begin{eqnarray}
 J_{T-AP}={2e \alpha \over \hbar P_o}(1- \eta)M_{\text
{s}}t(H + 4\pi M_{\text {eff}}). \; \label{eq2'}
\end{eqnarray}
$J_{T-P}$ and $J_{T-AP}$ cross and become equal at the field $4 \pi
M_{\text{eff}}/\eta$. However, hysteresis appears at a smaller
field, $H^*$, corresponding to the lowest field at which the AP/PS
region appears in the phase diagram. An interesting feature of the
perpendicular field phase diagram is that the hysteresis is
associated with the angular dependence of the spin-transfer torque
\cite{LiHeZhang2005}; the fact that the torque is larger in the AP
state ($\eta>0$). This is in contrast to conventional hysteresis in
magnets, which is associated with dipolar interactions or magnetic
anisotropy. Note that starting at large current and decreasing the
current, the model predicts an abrupt and large change in the angle
between the free and fixed layer at $J_{T-AP}$, which is detectable
as a step change in junction resistance. This suggests that a
measurement of the critical current on a decreasing current can be
used to determine the AP threshold current (Eq. 3) as a function of
magnetic field and sample structure. This is the approach we take in
analyzing our experiments. While we will focus on the high-field
behavior, we note for fields less than $4\pi M_{\text {eff}}$ the
layer magnetizations tilt into the film plane and the phase
boundaries depend on the in-plane magnetic anisotropy.

Hundreds of pillar junctions with submicron lateral dimension were
fabricated on a $1 \times 1 ~cm^2$ silicon wafer using a nanostencil
process \cite{Sun2002}. Stencil holes with different but accurate
lateral dimensions were opened up at the depth of $\sim 75 ~nm$, and
pillar junctions were deposited into those stencil holes through
metal evaporation. Junctions have the layer structure $\parallel 3
~nm ~Pt\mid 10 ~nm ~Cu\mid t ~Co\mid 10 ~nm ~Cu\mid 12 ~nm ~Co\mid
10 ~nm ~Cu\mid 3 ~nm ~Pt\mid 200 ~nm ~Cu\parallel$. During the
evaporation of the thin Co layer, we used a linear motion shutter to
vary $t$ from $1.8$ to $5.3 ~nm$ across the wafer. Junctions with
lateral dimensions $50 \times 50$, $50 \times 100$ and $70 \times
140 ~nm^2$ were studied in detail.

\begin{figure}[t]
\includegraphics[width=0.48\textwidth]{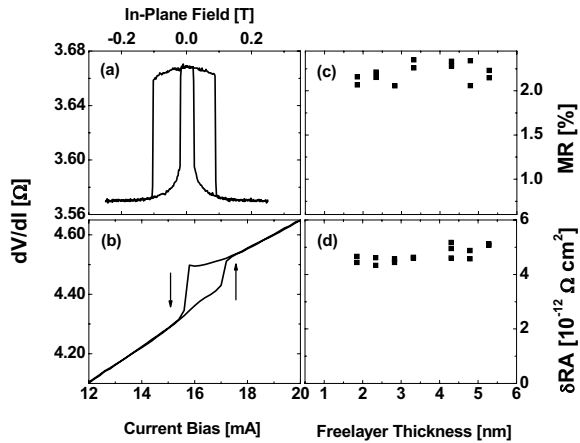}
\caption{(a) Zero dc current in-plane magnetoresistance hysteresis
loop for a $50 \times 50 ~nm^2$ junction with $t\simeq2.8 ~nm$. (b)
Positive current sweep hysteresis loop of the same junction with
perpendicular magnetic field set at $7 ~T$. (c) MR of all junctions
as the function of $t$. (d) Zero dc current in-plane $\delta R$
times lateral area $A$ for all junctions as the function of $t$. }
\label{Magnetoresistance}
\end{figure}

All measurements reported here were made at 4.2 K with a 4-point
geometry. Both dc resistance $V/I$ and differential resistance
$dV/dI$ were measured for each junction. A $0.2 ~mA$ modulation
current at $802 ~Hz$ was added to the dc bias. Junction resistances
were found to scale inversely with lateral areas. Positive current
is defined to be electron flow from the free (thin) Co layer to the
fixed (thick) Co layer.

The magnetoresistance (MR) was measured with the magnetic field
applied in the film plane. A typical MR hysteresis loop of a $50
\times 50 ~nm^2$ junction with $t\simeq 2.8 ~nm$ is shown in Fig.
\ref{Magnetoresistance}(a). The high resistance state corresponds to
an AP state and for fields greater than $0.1 ~T$ lead to a P state
of lower resistance. The magnetoresistance,
$MR=(R_{AP}-R_{P})/R_{P}=\delta R /R_{P}$, is $(2.2 \pm 0.2)\%$,
independent of the free layer thickness within the thickness range
investigated (Fig. 1(c)). This shows that the MR is due mainly to
spin-dependent scattering at Co/Cu interfaces. The MR area product,
$\delta R A$, where $A$ is the lateral area of the junction, is also
independent of thickness, which indicates that $\delta R$ is
inversely proportional to the junction area, as expected.

Current-voltage measurements were conducted with a magnetic field
applied perpendicular to the sample surface. Fig.
\ref{Magnetoresistance}(b) shows a current sweep hysteresis loop of
the same junction as in Fig. \ref{Magnetoresistance}(a) with a $7
~T$ applied field. When the current is swept up to a sufficiently
large value, $17.1 ~mA$ in this case, a step increase in resistance
is observed (indicated with an upward arrow in Fig.
\ref{Magnetoresistance}(b)). We interpret this step as indicating
the current at which the junction switches into an AP state, as
reported in Ref. \cite{Ozyilmaz2003}. For decreasing current there
is a step down in differential resistance at $15.7 ~mA$, which, as
discussed, we associate with the linear instability threshold given
in Eq. \ref{eq2'}, $J_{T-AP}$. The current hysteresis is about $1.4
~mA$ at this field. The majority of junctions show hysteresis in
current sweep measurements for fields larger than $3 ~T$ and steps
in both $dV/dI$ and $V/I$ \cite{footnote1}. The sharp step in $V/I$
has been used to determine the critical current for all junctions.

\begin{figure}[t]
\includegraphics[width=0.48\textwidth]{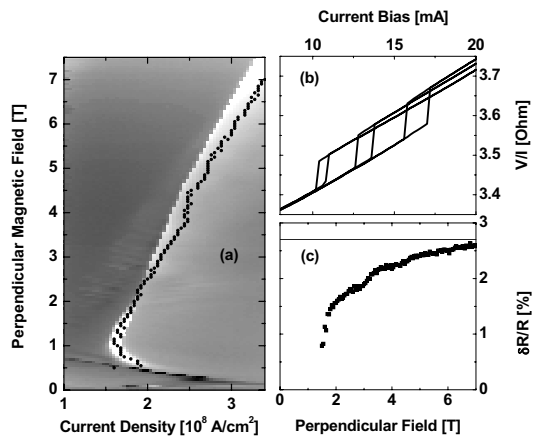}
\caption{(a) Contour plot of $dV/dI$ minus a linear background as a
function of both current density and magnetic field perpendicular to
sample surface for decreasing current. The same junction as in Fig.
2 (a,b). ``$\cdot$" data points: The corresponding step in $V/I$ for
increasing current. (b) $V/I$ \textit{vs} current bias hysteresis
loops with fields set at $3$, $5$ and $7 ~T$. (c) $\delta R/R$ at
the critical current as the function of field. Solid line: Zero dc
current in-plane MR. } \label{Switching}
\end{figure}

A contour plot of $dV/dI$ as a function of current and perpendicular
magnetic field shows the variation of the critical current with the
applied perpendicular field. In order to emphasize the change of
resistance on top of the background, which is associated with Joule
heating, we have plotted $dV/dI$ minus a linear background versus
current density. Fig. \ref{Switching}(a) shows data for decreasing
current on the same junction as in Fig \ref{Magnetoresistance}(a),
(b). The brighter the color in the contour plot, the larger the
junction differential resistance. The boundary between the bright
and dark region is an abrupt step in differential resistance. The
corresponding position of the abrupt step for increasing current is
illustrated with black dots in Fig. \ref{Switching}(a). We note that
this boundary is not as straight as the current sweep down boundary.
This is the case for all junctions, which we believe has its origin
in the fact that magnetization precession and possibly spatially
non-uniform spin-wave modes of the free layer, are excited at
currents just below this switching boundary (see Fig.
\ref{Numerical}). Above the demagnetization field ($\sim1.5 ~T$),
the threshold current increases with the applied magnetic field,
consistent with Eq. \ref{eq2'}. Fig. \ref{Switching}(b) shows the
steps in $V/I$ at $3$, $5$ and $7 ~T$. The hysteresis decreases with
decreasing field, and vanishes at $H^{*}$. $H^*$ is nearly
independent of thickness, and is $2.5\pm 0.3 ~T$ for most junctions
\cite{footnote1}. The change in DC resistance $\delta R/R$ at the
threshold current as a function of field is shown in Fig.
\ref{Switching}(c). The solid line at $2.7 ~\%$ is the in-plane MR.
$\delta R/R$ increases with increasing field and asymptotically
approaches the in-plane MR. In the macrospin model, more abrupt
changes in magnetization state occur at higher magnetic field, due
to the increasing importance of the angular dependence of the spin
transfer torque, parameterized by $\eta$ in our discussion. For
example, for fields greater than $4 \pi M_{\text{eff}}/ \eta$, the
model predicts switching between AP and P states at $J_{T-AP}$ and
thus the full MR. Fig. \ref{Switching}(b) also shows that there is
slight increase in the resistance at high current as the field
increases.

\begin{figure}[t]
\includegraphics[width=0.48\textwidth]{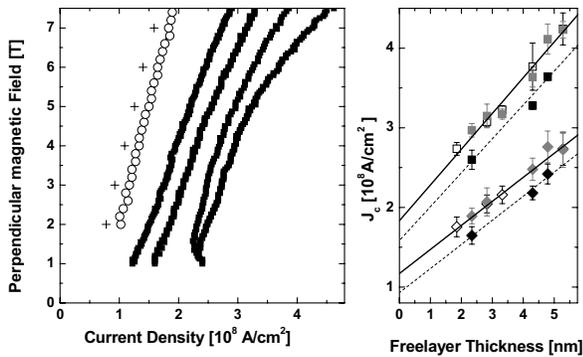}
\caption{(a) Solid points: critical-current density \textit{vs}
perpendicular magnetic field of 4 of the $50 ~nm$ series junctions
with $t=1.9, ~2.8, ~4.3, ~5.3 ~nm$ (from left to right). Open
circles: critical current densities of all $50 ~nm$ series junctions
extrapolated to zero $t$. Crosses: those of $70 \times 140~nm^2$
junctions. (b) Critical current densities as a function of free
layer thickness. Squares: 7 T. Diamonds: 3 T. Open, gray and black
data points are the ones of $50 \times 50$, $50 \times 100$ and $70
\times 140~nm^2$ junctions respectively. Solid lines: linear fits of
critical current densities \textit{vs} $t$ of $50 ~nm$ series.
Dashed lines: those of $70 \times 140~nm^2$ junctions.
}\label{ThicknessDependence}
\end{figure}

About 20 junctions with different $t$ and lateral dimensions were
measured. The critical-current densities $J_{c}$ as a function of
magnetic field for all junctions were measured and analyzed, and 4
of the junctions, representing different free-layer thicknesses, are
plotted in Fig. \ref{ThicknessDependence}(a). From the figure, it is
clear that the critical current increases with thickness at fixed
magnetic field. The critical current density at a fixed field for
all junctions is plotted in Fig. \ref{ThicknessDependence}(b) as a
function of a free-layer thickness, with different symbols (open,
gray, and black) denoting the three studied junction sizes, as shown
in the caption. The critical current density increases linearly with
thickness, with a slope that increases with increasing field. This
linear dependence is similar to the results of \cite{Albert2002}.

One interesting result from Fig. \ref{ThicknessDependence}(b) is
that the critical current density $J_{c}$ also shows an observable
dependence on lateral dimension. In our data, critical current
densities of $70 \times 140~nm^2$ junctions are generally lower than
those of the $50 \times 50$ and $50 \times 100 ~nm^2$ junctions
(denoted the $50 ~nm$ series for ease of reference). As a result,
$50 ~nm$ series (solid lines) and $70 \times 140~nm^2$ junctions
(dashed lines) were fit separately. In order to show the trend
within the same series, 4 typical switching boundaries only from the
$50 ~nm$ series were plotted in Fig. \ref{ThicknessDependence}(a).
From Fig. \ref{ThicknessDependence}(b), the $J_{c}$ shift of the $70
\times 140~nm^2$ junctions from that of the $50 ~nm$ series is
independent of the free-layer thickness, as the linear fits of the
two different lateral series are parallel to each other.

We find that the intercepts of $J_{c}$ \textit{vs} $t$ at different
fields are not zero, as seen from the linear fits in Fig.
\ref{ThicknessDependence}(b). The intercepts, denoted by $J_{c0}$,
of the $50 ~nm$ series are plotted as circles in Fig.
\ref{ThicknessDependence}(a), which is the ``switching boundary in
the limit of \emph{zero free-layer thickness}." The corresponding
boundary of the $70 \times 140~nm^2$ junctions is plotted as the
crosses in the same figure. $J_{c0}$ is field dependent, which shows
that $dJ_{c0}/dH$ is nonzero. Because of the curvature of $J_{c}$
\textit{vs} $H$ at high field for thicker free layers, $dJ_{c}/dH$
becomes field dependent, and $dJ_{c}/dH$ \textit{vs} $t$ turns out
to be nonlinear. Therefore, the extrapolation of $dJ_{c}/dH$ to zero
thickness becomes ambiguous. Interestingly, $J_{c}$ at a fixed field
shows a linear dependence on thickness despite the curvature (Fig.
\ref{ThicknessDependence}). This linear dependence unambiguously
gives a finite intercept of $J_{c}$ at zero free-layer thickness.
Note that although the curvature of the measured switching
thresholds increases at high fields for thicker free layers, the
slope of $J_{c0}$ \textit{vs} $H$ is constant from 2 to $7 ~T$.

$J_{c0}$ is affected by the lateral dimension, since there is a
shift between the boundaries of the two series. However, these two
boundaries are parallel to each other, which shows that $dJ_{c0}/dH$
is independent of the lateral dimension. $dJ_{c0}/dH=(1.5 \pm 0.3)
\times 10^{7} ~A/cm^{2} T$ for both \emph{zero free-layer thickness}
boundaries, while Eq. 3 predicts $dJ_{c0}/dH={2e \alpha \over \hbar
P_o}(1-\eta)M_{\text {s}}t|_{t=0}=0$.

The lateral dimension of the pillars affects the critical currents,
in that a constant shift of the critical-current densities between
the $50 ~nm$ series and $70 \times 140 ~nm^2$ was observed.
Micromagnetic analysis of thin disks \cite{Otani2004, McMichael2005}
shows that the principle mode of the free layer is influenced by the
lateral dimension and aspect ratio. As the result, the mode
frequency depends on the lateral dimension. Since the critical
current is proportional to the mode frequency, the change in the
normal mode frequency is expected to shift the intercept of $J_{c}$
\textit{vs} $H$, while not changing $dJ_{c0}/dH$. Experimentally
$dJ_{c0}/dH$ is not found to depend on lateral dimension.
Furthermore, in contrast to the prediction of Eq. \ref{eq2'}, the
threshold boundaries of most junctions do not extrapolate to $-4\pi
M_\text{eff}$ at zero current. Similarly, this may also be
associated with the shift of the normal mode precessional frequency.

We note that Brataas \textit{et al.} has recently computed the
critical current for the excitation of non-uniform spin-wave modes
\cite{Brataas2006}. The critical current density, considering the
effect of the non-uniform modes, was found to depend linearly on
magnetic-layer thickness --provided the excitation of such modes is
opposed by bulk Gilbert damping.

The magnetic anisotropic field associated with the easy plane
anisotropy is given as $4\pi M_{\text{eff}}$. Based on an FMR study
on extended Co films \cite{Beaujour2006}, $4\pi M_{\text{eff}}$ is a
function of Co thickness. In the thickness range studied here, $4\pi
M_{\text{eff}}$ changes as $1/t$, and it increases only by $0.2 ~T$
when Co thickness decreases from 5.3 to 1.9 nm. So anisotropy is
also a minor effect, which cannot be the main contribution to the
non-zero $dJ_{c0}/dH$.

There are two possible explanations for the observation of nonzero
$dJ_{c0}/dH$. The first is associated with an interface contribution
to the magnetization damping. Recently, Tserkovnyak \textit{et al.}
\cite{Tserkovnyak2005} employed a scattering theory approach to
characterize this contribution to the damping for a thin
ferromagnetic layer in contact with normal metals. They consider the
spin current into adjacent normal layers (N) when there is
magnetization dynamics. When such a spin current is dissipated by
spin-flip scattering in the N layers, it generates additional
damping. We can estimate this additional contribution to the damping
from the Tserkovnyak \textit{et al.} theory, assuming that the Co
layer is surrounded by perfect spin sinks, one of which is the fixed
magnetic layer, and the other of which is a Pt layer, separated by
10 nm of Cu from the thin Co layer. The additional spin-pumping
contribution to the damping is $\alpha ' \simeq 1.7 \times 10^{-2}
~nm/t$, and is consistent with our recent FMR studies on similar
structures \cite{Beaujour2006}. As a result the net damping is
$\alpha=\alpha_{o}+\alpha '$, where $\alpha_{o}$ is the bulk
damping. With an interface contribution to the damping, the
threshold current goes to a finite value in the zero thickness
limit. Quantitatively, taking $\eta=0.3$ and $P_o=0.3$, we find
$dJ_{c0}/dH=1.4 \times 10^{7} ~A/cm^{2} T$, which is to be compared
to our experimental result of $1.5 \times 10^{7} ~A/cm^{2} T$. This
quantitative agreement suggests interfacial damping plays a
significant role in determining the critical current density in
spin-transfer devices.

An alternative explanation for our results is due to Zhang and Levy
\cite{Zhang2004, Zhang2005}. In their model, the transverse
component of spin decays on a length scale of $\lambda_{J}$ in the
ferromagnet, which they find is about 3 nm for Co. As a result, for
layers of order and less than $\lambda_{J}$, the efficiency of the
spin-transfer torque decreases --as the transverse component of
angular momentum is not fully transferred to the thin magnetic
layer. $dJ_{T-AP}/dH$ decreases to a $t=0$ limit of ${2e \alpha_o
\over \hbar P_o }\lambda_{J} (1 -\eta)M_{s}=1.0 \times 10^{7}
~A/cm^{2} T$, with $\alpha_o=0.01$ \cite{Beaujour2006}. This is also
close to what we observe in experiment.

In summary, the phase diagrams for spin-transfer-induced magnetic
excitations are experimentally determined in perpendicular magnetic
field as a function of free-layer thickness. Based on the macrospin
model (Fig. \ref{Numerical}) and the observed hysteresis, we
estimate the ratio of the torque in the AP state to that in the P
state, due to its angular dependence, to be approximately 2 for a
constant current. Further, we have found that the critical-current
density for spin-transfer excitations is a linear function of
free-layer thickness that extrapolates to a finite critical current
at zero free-layer thickness. We have highlighted the role of the
spin-pumping contribution to the damping, which can quantitatively
explain our results. An implication of these results is that
reducing the thickness of the magnetic layers permits a reduction of
the critical-current density only to a lower limit set by interface
effects.

We thank P. M. Levy for many useful discussions of this work. This
research is supported by NSF-DMR-0405620 and by ONR N0014-02-1-0995.


\begin{thebibliography}{18}
\bibitem{footnote1} One exception is that hysteresis was not found on a $50 \times 50 ~nm^2$ junction with a $5.3 ~nm$ free layer. This could be due
to the strong coupling between the magnetic layers when free layer
is sufficiently thick.
\end{thebibliography}
\end{document}